\documentclass[12pt, letterpaper]{article}
\usepackage[margin=1in]{geometry}
\usepackage[utf8]{inputenc}
\usepackage{latexsym}
\usepackage{changepage}  
\usepackage{hyperref}
\usepackage{csvsimple}
\usepackage{titlesec}
\titlelabel{\thetitle.\quad}

\makeatletter
\renewcommand{\@seccntformat}[1]{~\csname the#1\endcsname{}.}
\makeatother

\newcommand{\Section}{\section}

\makeatletter
\csvset{
  my hack/.style={
    late after line=\\\hline,
    late after last line=\csv@tablefoot\end{tabular}\csv@posttable},
}
\makeatother

\begin{document}

\begin{center}
    
    \huge\textbf{An experimental study of algorithms for obtaining a singly connected subgraph}   \\[20pt]
    
    \Large{Ahmed Zahloote, Al-hasan Saleh, Ayman Ghanem, Hiba Hasan, Asem Dreibaty, Ali Abodaraa, Nermeen Suleiman,Nour Naameh,Ali Ibrahim, Zeinab mahfoud }\\

    \rule{\textwidth}{0.5pt}
    \begin{abstract}
    \normalsize{\noindent A directed graph $G = (V,E)$ is singly connected if for any two vertices $v,u \in V$, the directed graph $G$ contains at most one simple path from $v$ to $u$. In this paper, we study different algorithms to find a feasible but necessarily optimal solution to the following problem. Given a directed acyclic graph $G=(V,E)$, find a subset $H \subseteq E$ of minimum size such that the subgraph $(V,E \setminus H)$ is singly connected. Moreover, we prove that this problem can be solved in polynomial time for a special kind of directed graphs.}
    
    \end{abstract}
    \textbf{\normalsize {keywords}\\} 
    \small{Depth First Search, Singly Connected Graph, cross edges, forward edges, backward edges, Sources, Sinks}
    \rule{\textwidth}{0.4pt}
\end{center} 
\Section{Introduction}

A directed graph $G = (V,E)$ is singly connected if for any two vertices $v,u \in V$, the directed graph $G$ contains at most one simple path from $v$ to $u$ \cite{CormenLeisersonRivest1991,CormenLeisersonRivest2009}. Buchsbaum and Carlisle \cite{BuchsbaumCarlisle}, Khuller \cite{Khuller1999,Khuller2000}, and Karlin \cite{Karlin1995} provided algorithms to check whether an input directed graph is singly connected or not in $O(n^{2})$ time. In $2015$, Dietzfelbinger and Jaberi \cite{DietzfelbingerJaberi2015} presented a better algorithm for testing whether a directed graph is singly connected or not.
Moreover, they proved that the following problem is NP-hard. Given a directed graph $G=(V,E)$, find a subset $H \subseteq E$ of minimum size such that the subgraph $(V,E \setminus H)$ is singly connected. 
In this paper, we prove that this problem can be solved in polynomial time for a special kind of graphs. Furthermore,
we study different algorithms to find a feasible but necessarily optimal solution to this problem when the input graph has no cycles.

\section{An optimal solution for special kind of graphs}
In this section we study a special case when the input graph $G=(V,E)$ has the following properties: 
\begin{itemize}
\item $G$ has no cycles
\item the number of sources in G is one
\item the number of verices is equal to the number of edges in $G$
\end{itemize}

The following Theorem shows that removing only one edge leaves a singly connected subgraph. This means we can find an optimal solution which contains only one edge.

\textbf{Theorem $1$}: Let $G=(V,E)$ be a directed acyclic graph such that $|E|=|V|$ and the number of sources in $G$ is one. Then an optimal solution can be found in polynomial time.

\textbf{proof.} Let $h$ be the source in $G$. Then all the veritices in $V$ are reachable from the vertex $h$ \cite{SedgewickWayne2011}. Therefore, performing depth first search algorithm starting from vertex $h$ produces exactly $|V|-1$ tree edges and one non-tree edge which is a cross or forward edge. Moreover, these tree edges form a tree called dfs tree. This tree is a subgraph of $G$ and for any two vertices $v,u \in V$, there is at most one simple path from $v$ to $u$ in this tree. Notice that the non-tree edge produced by depth first search is an optimal solution.

The following pseudocode (DFS\_once) shows an algorithm that is able to compute an optimal solution when the input graph is a directed acyclic graph with one source and the number of edges is equal to the number of vertices.

DFS\_once(DirectedGraph):\\
1. Let $h$ be the source source in $G$\\
2. tree $\longleftarrow$ DFS(DirectedGraph)\\
3. for each edge in tree:\\
\{\\
    \begin{adjustwidth}{1cm}{}
    3.1 if(edge.class.is\_equal\_to("cross") or edge.class.is\_equal\_to("forward"))\\
    \{
        \begin{adjustwidth}{1cm}{}
         3.1.1 cross\_edge $\longleftarrow$ edge
        \end{adjustwidth}
    \}
    \end{adjustwidth}
\}\\    
4. DirectedGraph.remove\_edges\_from(cross\_edge)\\
5. return DirectedGraph

The correctness of this algorithm follows from Theorem $1$ and the following theorem shows that DFS\_once algorithm runs in linear time.\\
\textbf{Theorem $2$.} DFS\_once algorithm runs in $O(n)$ time.\\
\textbf{proof.} DFS runs in linear time and the number of edges is $n$.

\section{Algorithms for directed graphs that has no cycles}
In this section we describe different algorithms for obtaining a singly connected subgraph when the input graph contains no cycles.

\subsection{Algorithm 1}
The first algorithm is about doing DFS (Depth First Search) on each node with indegree zero (called a source node) and classify each edge during the search then we combine the edges classified as cross or forward edges and delete them from graph, here is the pseudo code:\\

DFS\_From\_Sources(DirectedGraph):\\
\{\\
1. non\_tree\_edges[ ] $\longleftarrow$ [ ] \\
2. sources[ ] $\longleftarrow$ [ ] \\
3. for each node in Directedgraph do \\
\{  
\begin{adjustwidth}{1cm}{}
    3.1 deg $\longleftarrow$ node.cal\-degree(node)\\
    3.2 if(deg.is\_equal\_to(0)\\
    \{
    \begin{adjustwidth}{1cm}{}
    3.2.1 sources.add(node)
    \end{adjustwidth}
    \}
\end{adjustwidth}    
\}\\
4. for each source in sources do\\
\{  
\begin{adjustwidth}{1cm}{}
    4.1 tree$\longleftarrow$ DFS(Source) /*this classifies edges of tree  to cross,forward and tree edges*/\\
    4.2 for each edge in tree do
    \{\begin{adjustwidth}{1cm}{}
    4.2.1 if(edge.class.is\_equal\_to("cross") or edge.class.is\_equal\_to("forward"))\\
    \{
    \begin{adjustwidth}{1cm}{}
    4.2.1.1 non\_tree\_edges.add(edge)
    \end{adjustwidth}  
     \}
    \end{adjustwidth}
\end{adjustwidth}
\}\\
5. DirectedGraph.remove\_edges\_from(non\_tree\_edges[]) \\
6. return DirectedFraph\\
\}

\textbf{Theorem $3$.} The running time of DFS\_From\_Sources algorithm is $O(n(n+m)$ \\
\textbf{Proof.} Calculating the indegree for all vertices requires $O(n+m)$ time. Furthermore, DFS runs in linear time and edges can be classified using dfs in linear time.

 \textbf{Theorem 4:} The first algorithm (DFS\_from\_sources) returns a singly connected subgraph.\\
 \textbf{proof.} Let $H$ be the set of non-tree edges computed in DFS\_from\_sources algorithm. Suppose that the subgraph $(v,E\setminus H)$ is not singly connected. Then there are two vertices $u,v \in V$ such that the subgraph $(V,E\setminus H)$ contains two edge-disjoint paths $\emph{P}$,$\acute{P}$ from $u$ to $v$ in $G$
Consequently, since the subgraph $(V,E\setminus H)$ is a directed acyclic graph there is a source $s \in V $ such that all the vertices and edges of the paths P,$\acute{P}$ are reachable from the source $s$. Notice that the DFS tree rooted at $s$ can't have two tree edges entering $u$.

\subsection{Algorithm 2}
The second algorithm is similar to the first one but the difference is that the execution of DFS will depend on the number of sources and sinks so if there are more sinks than sources  about reversing the direction of all the edges and then executing DFS (Depth First Search) on each node with outdegree of the node equal to zero (a sink node) and classify each edge during the search then we combine the edges classified as cross or forward edges and delete corresponding original edges from the graph, here is the pseudo code:\\

DFS\_From\_Sources$|$Sinks(DirectedGraph):\\
\{\\
1. non\_tree\_edges[ ] $\longleftarrow$ [ ] \\
2. sinks[ ] $\longleftarrow$ [ ]  \\
3. sources[ ] $\longleftarrow$ [ ]  \\
4. for each node in Directedgraph do \\
\{  
\begin{adjustwidth}{1cm}{}
    4.1 indeg $\longleftarrow$ node.cal\-degree()\\
    4.2 outdeg $\longleftarrow$ node.cal\_outdegree()\\
    4.3 if(indeg.is\_equal\_to(0)\\
    \{
    \begin{adjustwidth}{1cm}{}
    4.3.1 sources.add(node)
    \end{adjustwidth}
    \}
    4.4 if(outdeg.is\_equal\_to(0)\\
    \{
    \begin{adjustwidth}{1cm}{}
    4.4.1 sinks.add(node)
    \end{adjustwidth}
    \}
\end{adjustwidth}    
\}\\
5. max $\longleftarrow$ max(sources,sinks)\\
6. for each node in max do\\
\{  
\begin{adjustwidth}{1cm}{}
    6.1 tree$\longleftarrow$ DFS(node) /*this classifies edges of tree  to cross,forward and tree edges*/\\
    6.2 for each edge in tree do
    \{\begin{adjustwidth}{1cm}{}
    6.2.1 if(edge.class.is\_equal\_to("cross") or edge.class.is\_equal\_to("forward"))\\
    \{
    \begin{adjustwidth}{1cm}{}
    6.2.1.1 non\_tree\_edges.add(edge)
    \end{adjustwidth}  
     \}
    \end{adjustwidth}
\end{adjustwidth}
\}\\
7. DirectedGraph.remove\_edges\_from(non\_tree\_edges[]) \\
8. DirectedGraph.reverse\_edges\\
9. return DirectedGraph\\
\}

\textbf{Theorem $5$.} Algorithm DFS\_From\_Sources$|$Sinks runs in O(max$\emph{ \{i1,i2} \} $.($\emph {n+m}$)) time where\emph{i1} is the number of sources and \emph{i2} is the number of sinks, the second algorithm pretty similar to the first one but the only difference is that edges may be reversed based on the biggest number between sources and sinks(if the sinks are more than the sources)

The correctness of the second algorithm (DFS from max(sinks,sources))follows from theorem $4$ and the following observation, G is singly connected if and only if $G^R$ is singly connected.

\subsection{Algorithm 3}
The third algorithm calculates dfs tree rooted at each source. The edges of these dfs trees form a subgraph which is not necessarily singly connected. Then the algorithm performs the first algorithm on the obtained subgraph, here is the pseudo code:

DFS\_on\_tree\_edges(DirectedGraph):\\
\{\\
1. non\_tree\_edges[ ] $\longleftarrow$ [ ]  \\
2. sinks[ ] $\longleftarrow$ [ ]  \\    
3. sources[ ] $\longleftarrow$ [ ] \\
4. for each node in Directedgraph do \\
\{  
\begin{adjustwidth}{1cm}{}
    4.1 indeg $\longleftarrow$ node.cal\-degree()\\
    4.2 outdeg $\longleftarrow$ node.cal\_outdegree()\\
    4.3 if(indeg.is\_equal\_to(0)\\
    \{
    \begin{adjustwidth}{1cm}{}
    4.3.1 sources.add(node)
    \end{adjustwidth}
    \}
\end{adjustwidth}    
\}\\
5. tree\_edges\_set$\longleftarrow$ \{\}\\
6. for each source in sources do\\
\{
\begin{adjustwidth}{1cm}{}
    6.1 tree$\longleftarrow$ DFS(Source)\\
    6.2 for each edge in tree do:\\
    \{\\
    \begin{adjustwidth}{1cm}{}
    6.2.1 tree\_edges\_set $\longleftarrow$ tree\_edges\_set + edge
    \end{adjustwidth}
    \}
\end{adjustwidth}
\}\\
7. DirectedGraph.remove\_edges\_from(tree\_edges\_set)\\
8. for each source in sources do\\
\{
\begin{adjustwidth}{1cm}{}
8.1 tree $\longleftarrow$ DFS(Source)\\
8.2 for each edge in tree do:\\
\{
    \begin{adjustwidth}{1cm}{}
    8.2.1 if(edge.class.is\_equal\_to("cross") or edge.class.is\_equal\_to("forward"))\\
    \{
        \begin{adjustwidth}{1cm}{}
         8.2.1.1  non\_tree\_edges.add(edge)
        \end{adjustwidth}  
     \}
    \end{adjustwidth}
    \}
    
\end{adjustwidth}
\}\\
9. DirectedGraph.remove\_edges\_from(non\_tree\_edges)\\
10. return DirectedGraph\\

\textbf{Theorem $6$} The running time of the algorithm (DFS\_on\_tree\_edges) is O($\emph{i}$($\emph{n+m}$)).

Notice that the tree edges calculated by performing dfs on each source form a subgraph but necessarily singly connected. But when we perform the first algorithm on this subgraph we can obtain a singly connected subgraph by removing cross and forward edges. Therefore, the correctness of algorithm DFS\_on\_tree\_edges follows from Theorem $4$.

\subsection{Algorithm 4:}
Algorithm 4  executes DFS search on every medial node (medial means that it's not a source or a sink), after that we perform algorithm1 on the resulting subgraph, the pseudo code:

DFS\_from\_medials(DirectedGraph):\\
1. non\_tree\_edges[ ] $\longleftarrow$ [ ]  \\
2. medials[ ] $\longleftarrow$ [ ]  \\    
3. for each node in Directedgraph do \\
\{  
\begin{adjustwidth}{1cm}{}
    3.1 indeg $\longleftarrow$ node.cal\-degree()\\
    3.2 outdeg $\longleftarrow$ node.cal\_outdegree()\\
    3.3 if(indeg.is\_not\_equal\_to(0) \& outdeg.is\_not\_equal\_to(0) )\\
    \{
    \begin{adjustwidth}{1cm}{}
    3.3.1 medials.add(node)
    \end{adjustwidth}
    \}
\end{adjustwidth}    
\}\\
4. for each medial in medials do\\
\{  
\begin{adjustwidth}{1cm}{}
    4.1 tree$\longleftarrow$ DFS(medial) 
    4.2 for each edge in tree do
    \{\begin{adjustwidth}{1cm}{}
    4.2.1 if(edge.class.is\_equal\_to("cross") or edge.class.is\_equal\_to("forward"))\\
    \{
    \begin{adjustwidth}{1cm}{}
    4.2.1.1 non\_tree\_edges.add(edge)
    \end{adjustwidth}  
     \}
    \end{adjustwidth}
\end{adjustwidth}
\}\\
5. TempGraph $\longleftarrow$ Graph(DirectedGraph.nodes(), non\_tree\_edges)\\
6. execute Algorithm 1 on TempGraph\\
7. return TempGraph\\
\}

\textbf{Theorem $7$}. algorithm DFS\_from\_medials runs in $O(n(n++m))$ time.

\section{Study Results:}
the following table contains the results for the execution of the algorithms on three graphs two of them were taken from SNAP,
the first one is p2p-Gnutella04 \cite{LeskovecKrevl2014,Leskovec2007,Ripeanu2002} and the second one is soc-Epinions1\cite{LeskovecKrevl2014,Richardson2003} (we removed all cycles from p2p-Gnutella04 and soc-Epinions1),  the third one it's a simple random acyclic graph that we built to try the algorithms on small inputs, the cpu used to execute code on the data is Intel core i7 11370h and 16GBs of memory, here are the results:

\csvautotabular[my hack]{results.csv}

\section*{Acknowledgements.}
We gratefully give all appreciation and thanks to Dr. Raed Jaberi for teaching and lecturing us approximation algorithms and advanced algorithms.


\addcontentsline{toc}{section}{References}
\begin{thebibliography}{1}
\bibitem{BuchsbaumCarlisle} A.L. Buchsbaum, M.C. Carlisle, Determining uni-connectivity in directed graphs, Information Processing Letters 48(1)(1993) 9--12.
\bibitem{CormenLeisersonRivest1991} T.H. Cormen, C.E. Leiserson, R.L. Rivest, Introduction to Algorithms,
The MIT Electrical Engineering and Computer Science Series, MIT Press, Cambridge, MA, 1991.
\bibitem{CormenLeisersonRivest2009} T.H. Cormen, C.E. Leiserson, R.L. Rivest, C. Stein, Introduction to Algorithms, MIT Press, third edition, 2009.
\bibitem{DietzfelbingerJaberi2015}Martin Dietzfelbinger, Raed Jaberi:
On testing single connectedness in directed graphs and some related problems. Inf. Process. Lett. 115(9): 684--688 (2015)
\bibitem{Khuller1999} S. Khuller, An O($ |V |^{2}$) algorithm for single connectedness, Information Processing Letters, 72(3--4)(1999) 105--107.
\bibitem{Khuller2000} S. Khuller, Addendum to "An O($ |V |^{2}$) algorithm for single connectedness", Information Processing Letters, 74(5--6) (2000) 263.
\bibitem{Karlin1995} A. Karlin, Solution to homework 7, CS 421, Winter 1995, Department of Computer Science, University of Washington, 1995.
\bibitem{LeskovecKrevl2014}Jure Leskovec and Andrej Krevl,https://snap.stanford.edu/data/,2014

 \bibitem{Leskovec2007}   J. Leskovec, J. Kleinberg and C. Faloutsos. Graph Evolution: Densification and Shrinking Diameters. ACM Transactions on Knowledge Discovery from Data (ACM TKDD), 1(1), 2007.
\bibitem{Ripeanu2002}    M. Ripeanu and I. Foster and A. Iamnitchi. Mapping the Gnutella Network: Properties of Large-Scale Peer-to-Peer Systems and Implications for System Design. IEEE Internet Computing Journal, 2002.

\bibitem{Richardson2003} M. Richardson and R. Agrawal and P. Domingos. Trust Management for the Semantic Web. ISWC, 2003.
 \bibitem{SedgewickWayne2011} Robert Sedgewick, Kevin Wayne:
Algorithms, 4th Edition. Addison-Wesley 2011, ISBN 978-0-321-57351-3, pp. I-XII, 1-955
\end{thebibliography}
\end{document}